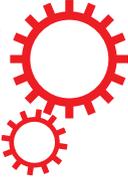

# SCIENTIFIC REPORTS

# The habitability of the Milky Way during the active phase of its central supermassive black hole

Amedeo Balbi [1] & Francesco Tombesi[1,2,3]

During the peak of their accretion phase, supermassive black holes in galactic cores are known to emit very high levels of ionizing radiation, becoming visible over intergalactic distances as quasars or active galactic nuclei (AGN). Here, we quantify the extent to which the activity of the supermassive black hole at the center of the Milky Way, known as Sagittarius A* (Sgr A*), may have affected the habitability of Earth-like planets in our Galaxy. We focus on the amount of atmospheric loss and on the possible biological damage suffered by planets exposed to X-ray and extreme ultraviolet (XUV) radiation produced during the peak of the active phase of Sgr A*. We find that terrestrial planets could lose a total atmospheric mass comparable to that of present day Earth even at large distances (~1 kiloparsec) from the galactic center. Furthermore, we find that the direct biological damage caused by Sgr A* to surface life on planets not properly screened by an atmosphere was probably significant during the AGN phase, possibly hindering the development of complex life within a few kiloparsecs from the galactic center.

The search for habitable planets beyond Earth is one of the major goals of contemporary astrophysics. This has motivated increased interest in quantifying the astrophysical factors that can make a planet unsuitable for life as we know it. It has long been recognized that prolonged exposition to ionizing radiation from space can hinder habitability by causing atmosphere loss and exposing planetary surfaces to biologically hazardous high-energy fluxes. Most work on the relation between habitability and exposure to ionizing radiation focused on exploring the effect of the activity of the host star on planetary systems, particularly for planets orbiting highly active stars such as M-dwarfs[1–5]. Other studies addressed the effect of ionizing events from transient high-energy sources such as supernovae and gamma ray bursts on their surroundings, including their potential impact on the history of life on Earth[6]. This led to the identification of regions in the Milky Way (a.k.a. Galactic Habitable Zone[7,8]) which might be less prone to such potentially life-threatening occurrences.

From the comparison with observational studies of the global statistical distribution of active galactic nuclei (AGN) with respect to their cosmological redshift and luminosity, we infer that Sgr A* has almost certainly been an important source of ionizing radiation during the peak of its active phase that has occured less than 8 Gyrs ago[9–11], and may have lasted between ~$10^7$ and $10^9$ years[12]. Previous studies on the relation between AGN and habitability have focused either on the effect of the activity of Sgr A* on the past environment of Earth[13] or, more recently, on the statistical modeling of the effect on planetary atmospheres of generic quasars populations in a cosmological context[14,15]. Here, we ask a different, more specific question: was the global habitability of the Milky Way affected by the past history of the supermassive black hole at its core? We address this issue by using plausible scenarios for the accretion history of Sgr A* and estimating the effect of its XUV emission on planetary atmospheres at various distances from the Galactic center. We then quantify the biological damage caused by the ionizing radiation produced by Sgr A* during its peak activity and whether this posed an increased hazard to surface life, partifcullarly for planets that lacked protection from a substantial gas envelope.

## Results

**Atmospheric loss.**    When XUV radiation falls upon a planetary atmosphere, it can cause its evaporation via hydrodynamic escape[16,17]. The mass loss rate depends on the XUV flux and on the bulk density of the planet. The efficiency of conversion of incident power into atmospheric loss is parameterized by a factor $\varepsilon$ (see Methods). The flux received by a planet at distance $D$ from the galactic center also depends on the attenuation of XUV radiation by the torus of material surrounding the supermassive black hole, parameterized by an optical depth $\tau$ (see Methods). We made the reasonable assumption that the AGN torus is co-aligned with the Galactic plane and considered two cases of $\tau \simeq 0$ (no attenuation) for high galactic latitude locations and $\tau \simeq 1$ (maximal attenuation) for locations close to the equatorial direction, respectively. We show in Fig. 1 our estimate of the total amount of atmospheric mass lost at the end of the AGN phase of Sgr A* by a rocky planet with the same density of the Earth as a function of the distance from the Galactic center.

[1]Dipartimento di Fisica, Università degli Studi di Roma "Tor Vergata", Via della Ricerca Scientifica, 00133, Roma, Italy. [2]Department of Astronomy, University of Maryland, College Park, MD 20742, USA. [3]NASA/Goddard Space Flight Center, Code 662, Greenbelt, MD 20771, USA. Correspondence and requests for materials should be addressed to A.B. (email: balbi@roma2.infn.it)

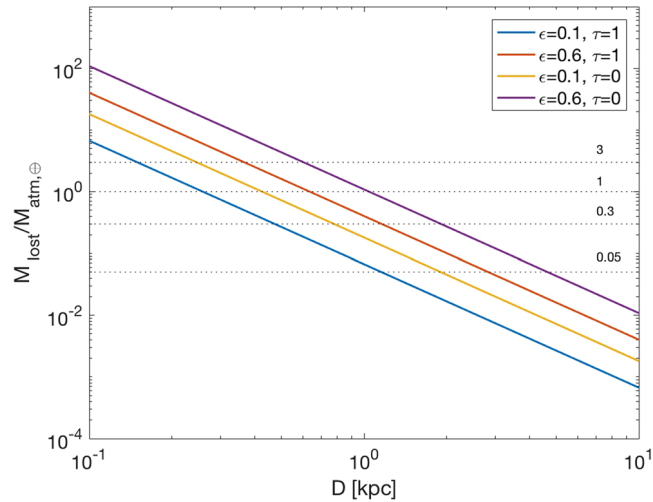

**Figure 1.** The total mass lost at the end of the AGN phase of Sgr A* by a terrestrial planet at distance D from the galactic center, in units of the atmosphere mass of present day Earth. Each curve was computed assuming a value for the efficiency of hydrodynamic escape of either $\epsilon = 0.1$ or $\epsilon = 0.6$. An optical depth $\tau \simeq 1$ corresponds to locations close to the galactic plane (maximum attenuation by the AGN torus) while $\tau = 0$ corresponds to high galactic latitudes (no attenuation).

Figure 2 shows how the amount of atmospheric mass lost depends on galactic location and on time, during the AGN phase of Sgr A*. We assumed that the radiative output of the supermassive black hole remained constant at ~10% of the Eddington luminosity during the active phase. This corresponds to a total duration of the active phase of ~5 × 10^8 yrs. A lower luminosity would result in a longer duration, and vice versa (see Methods). However, since the fluence would remain the same, the total atmospheric mass lost at the end of the AGN phase (Fig. 1) would not be affected. The distances from the galactic center corresponding to given values of total atmospheric mass lost are summarized in Table 1.

**Biological consequences.** Assessing a hazard threshold for life exposed to ionizing radiation is not straightforward. Lethal values in the literature are usually quoted as an absorbed dose, expressed in units of gray, equivalent to $10^4$ erg g$^{-1}$. This can be converted into a fluence if the mass absorption features of the biological material are known (often, liquid water is used as a proxy). However, the same total dose can result in smaller or even negligible biological damages if it is absorbed over a sufficiently long time, for example because cell repair mechanisms can come into effect. Then, ionizing radiation should exceed some critical flux (i.e. dose rate) in order to pose a significant hazard to living organisms. Furthermore, it is conceivable that extraterrestrial organisms might evolve their own radiation-resistance mechanisms under very different environmental conditions, making it difficult to predict a universal lethal flux threshold. We should also note that even radiation doses smaller than the lethal value could have relevant biological consequences by causing mutations or adding selective pressure to living organisms.

With these caveats in mind, we derived some quantitative indication of the possible biological damage on planetary surface life during the peak activity of Sgr A*. We did so by comparing the XUV flux at various distances from the galactic center to the absorbed dose that proved to be lethal for various terrestrial organisms. We adopted as reference a lethal dose of 1 gray for eukaryotes and multicellular life and of 100 gray for prokaryotes[18–21], corresponding to fluences ~5 × 10^5 erg cm$^{-2}$ and ~5 × 10^7 erg cm$^{-2}$, respectively[18,19]. We note, however, that the most radiation-resistant known terrestrial prokaryotes, such as *Deinococcus radiodurans*, can survive to acute doses of order $10^4$ Gray, or fluences[20,21] ~5 × 10^9 erg cm$^{-2}$. Since the lethal doses quoted in studies on the resistance of living organisms to radiations were absorbed over periods ranging from days to, at most, about a year, we decided conservatively to adopt a conventional reference value of 1 year to convert from fluence to flux. This results in critical fluxes of $1.6 \times 10^{-2}$ erg s$^{-1}$ cm$^{-2}$ for "complex" life and 1.6 erg s$^{-1}$ cm$^{-2}$ for prokaryotes. We then calculated the distance from Sgr A* where a planetary surface was exposed to such fluxes during peak activity. The results are summarized in Table 2.

**Discussion**

A robust conclusion of our study is that rocky planets in the Galactic bulge ($D \lesssim 1$ kpc) should have been exposed to enough XUV radiation during the AGN phase of Sgr A* to lose a significant fraction of their atmosphere. The total mass loss could be comparable to that of the atmosphere of present day Earth for distances $D \lesssim 0.5$ kpc even for locations close to the galactic plane, where the attenuation from the AGN torus is maximal, assuming values on the higher end of hydrodynamic efficiency. This is basically independent of specific assumptions on the accretion luminosity of the supermassive black hole. This effect may have catastrophically compromised the past and future habitability of planets in the Galactic bulge, unless their atmospheres were regenerated by some planetary processes (e.g. outgassing). As the rate of volcanism and outgassing for a terrestrial planet cannot be easily predicted from general models, we refrain here to incorporate such a source term in Equation 2, and we leave a quantitative estimation of this effect for future studies. However, we point out that our results can also be seen as a lower limit on the amount of atmospheric production needed for a terrestrial planet to outcompete loss and remain habitable after the end of the AGN phase.

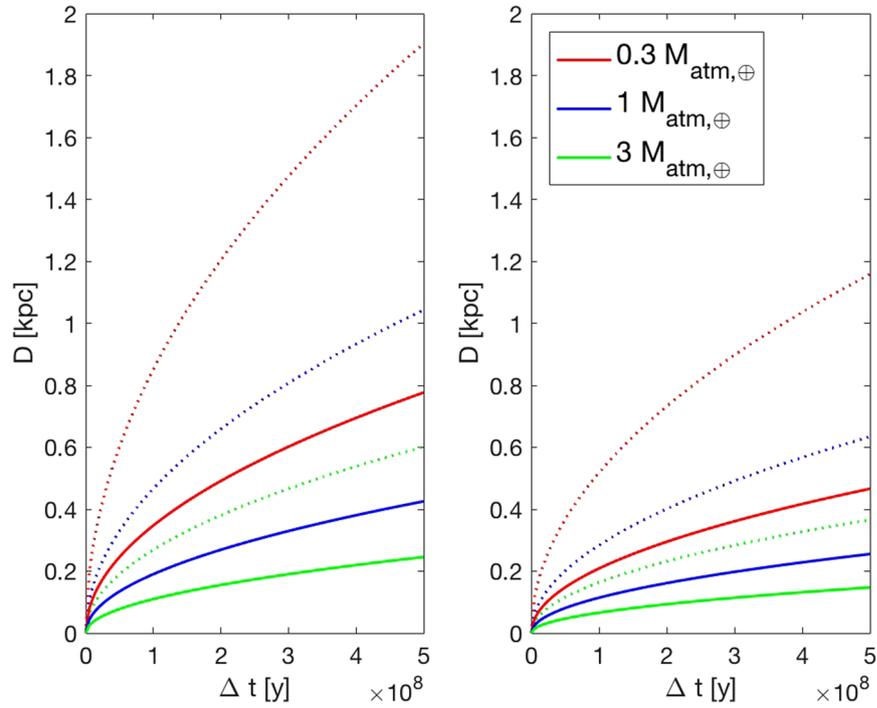

**Figure 2.** Atmospheric mass lost by a terrestrial planet as a function of time and galactic location, during the AGN phase of Sgr A*, assuming a constant radiative output ~10% $L_{Edd}$. The horizontal axis indicates the time passed from the beginning of the AGN phase, while the vertical axis is the distance of the planet from the Galactic center. The continuous curves represent a mass loss of 0.3 (red), 1 (blue) and 3 (green) in unit of present day Earth's atmosphere mass, assuming an efficiency factor for hydrodynamic escape of $\varepsilon = 0.1$, while the dotted lines are for $\varepsilon = 0.6$. The curves in the left panel were computed with no attenuation by the AGN torus ($\tau = 0$), corresponding to locations at high galactic latitudes, while the curves in the right panel are for maximum attenuation ($\tau \simeq 1$), corresponding to locations close to the galactic plane.

| | $0.05 M_{atm,\oplus}$ | $0.3 M_{atm,\oplus}$ | $1 M_{atm,\oplus}$ | $3 M_{atm,\oplus}$ |
|---|---|---|---|---|
| $\varepsilon = 0.6, \tau = 0$ | 4.66 | 1.90 | 1.04 | 0.60 |
| $\varepsilon = 0.1, \tau = 0$ | 1.90 | 0.78 | 0.43 | 0.25 |
| $\varepsilon = 0.6, \tau = 1$ | 2.84 | 1.16 | 0.63 | 0.37 |
| $\varepsilon = 0.1, \tau = 1$ | 1.16 | 0.47 | 0.26 | 0.15 |

**Table 1.** Distances in kpc from the galactic center corresponding to a given total mass loss at the end of the AGN phase of Sgr A*. Each column corresponds to a fraction of the atmosphere mass of present day Earth ($0.05 M_{atm,\oplus}$ is approximately the mass of present day Mars' atmosphere). Each line was computed for various combinations of the efficiency factor for hydrodynamic escape $\varepsilon$ and attenuation by the AGN torus $\tau$.

| | "complex" life | prokaryotes |
|---|---|---|
| $L = L_{Edd}, \tau = 0$ | 13.3 | 1.33 |
| $L = 10\% \ L_{Edd}, \tau = 0$ | 4.20 | 0.42 |
| $L = 1\% \ L_{Edd}, \tau = 0$ | 1.33 | 0.13 |
| $L = L_{Edd}, \tau = 1$ | 8.08 | 0.81 |
| $L = 10\% \ L_{Edd}, \tau = 1$ | 2.55 | 0.25 |
| $L = 1\% \ L_{Edd}, \tau = 0$ | 0.81 | 0.081 |

**Table 2.** Distances in kpc from the galactic center corresponding to high biological hazard for life on a fully exposed planetary surface, during the AGN phase of Sgr A*. The first column is the lethal distance for eukaryotes and multicellular organisms ("complex" life), the second is for prokaryotes. Each line was computed for different combinations of the supermassive black hole luminosity and attenuation by the AGN torus $\tau$.

A potential limitation of our study is the treatment of atmospheric mass loss through an energy-limited hydrodynamic escape model: this has the appeal of simplicity, since the result only depend on the total XUV flux, but it may not take fully into account the complexity of the interaction of the incoming radiation with the atmosphere. While we are confident from previous studies that this uncertainty is properly accounted for in the range of values of the efficiency parameter $\varepsilon$ we adopted, a full numerical treatment of radiative transfer, conduction and photochemical modeling of the atmosphere is beyond the scope of the present work. By pointing out the potential impact of the early AGN phase of the central black hole on galactic habitability, our work certainly encourages further investigations in this direction.

Another important conclusion of our study is that the high level of ionizing radiation falling on planetary surfaces lacking substantial protection from an atmosphere would have had dramatic biological consequences for land organisms. Assuming plausible values for the lethal absorbed dose of ionizing radiation, we found that the development of complex life was probably severely hindered during the AGN phase of Sgr A*, even at large distance from the galactic center (~1–10 kpc). This may not have prevented the appearance of life *per se*, since prokaryotes could have survived to higher fluxes. However, we point out that the biological effect of the ionizing radiation from Sgr A* would have been in addition to that of any other source of ionizing radiation, for example from the host star, and would add to the loss of a large fraction of the atmosphere. Even if some organisms might have survived by developing radiation-resistance or finding protected niches, the global effect on the biosphere would have been significant.

To sum up, our results imply that the inner region of the Milky Way might have remained unhabitable until the end of the AGN phase of its central black hole, and possibly thereafter. This has important consequences in assessing the likelihood of ancient life in the Galaxy, and should be taken into account in future studies of the Galactic habitable zone. It also suggests further investigations on the relation between supermassive black holes in galactic cores and planetary habitability.

## Methods

**The XUV emission of Sgr A*.**    In order to estimate the luminosity of Sgr A* during the peak of its activity we can first consider the Eddington luminosity defined as $L_{Edd} \simeq 1.26 \times 10^{38}(M_{BH}/M_\odot)$ erg s$^{-1}$. Using the best estimate for the mass of Sgr A*, $M_{BH} = (3.6 \pm 0.3) \times 10^6 \, M_\odot$ (ref.[22]), would result in a luminosity $L_{Edd} \simeq 4.5 \times 10^{44}$ erg s$^{-1}$.

Then, we can estimate the duration of the AGN phase considering the Salpeter time, defined as the time it takes to double the black hole mass during an Eddington accretion phase[14]. The relation between AGN luminosity $L$ and mass accretion rate $M_{acc}$ is $L = \eta \dot{M}_{acc} c^2$, where $\eta \simeq 0.1$ is the typical radiative efficiency. The rate of mass that eventually falls into the black hole is $\dot{M}_{BH} = (1 - \eta)\dot{M}_{acc}$. We can then estimate the Salpeter time as $\Delta t_S = M_{BH}/\dot{M}_{BH}$. Substituting the parameters using the previous equations, and assuming $L = L_{Edd}$, we obtain a mass-independent time-scale of $\Delta t_S \simeq 5 \times 10^7$ yrs.

We can consider the Eddington luminosity as an upper limit on the AGN radiative output and therefore the resultant Salpeter time as a lower limit on the timescale of the AGN phase. Considering the most typical luminosity value of ~10% $L_{Edd}$ for local Seyfert galaxies[23] the AGN phase would be 10 times longer, $\Delta t = 5 \times 10^8$ yrs. Considering an even lower luminosity regime of only ~1% $L_{Edd}$, the timescale would be 100 times longer than $\Delta t_S$. We consider the ~10% $L_{Edd}$ to be the most likely accretion rate during the AGN phase. However, we note that, since the total atmospheric mass lost by a planet depends on the integrated fluence ~$L\Delta t$ (as explained later), the final result at the end of the AGN phase will not depend on the assumed luminosity.

Finally, we need an estimate of the amount of XUV radiation emitted during the AGN phase. The AGN broad-band spectral energy distribution is well approximated by a power-law $F(\lambda) = A\lambda^a$, where the spectral index is typically[23] $\alpha \simeq -1$. We estimated the XUV luminosity as the integral $L_{XUV} = \int d\lambda F(\lambda)$ over the range $1.24 \times 10^{-3} \le \lambda \le 1.24 \times 10^2$ nm (that is, at energies 10–10$^6$ eV). This is roughly 70% of the total luminosity over the range 0.1–10$^6$ eV, consistent with population studies which show that the AGN power is mostly released in the XUV band[23,24].

**Attenuation of XUV radiation by the AGN torus and the insterstellar medium.**    The unified model of AGN[25] postulates that the accreting supermassive black hole is surrounded by an optically-thick torus at scales of about 1 pc. This would cause an attenuation $e^{-\tau}$ of the flux from the AGN: the optical depth $\tau$ of the torus is null along the polar direction and it reaches a maximum of $\tau \simeq 1$ at the equator[26]. Flux attenuation by the interstellar medium in the Galaxy is negligible with respect to the AGN torus because it is always optically thin and it may reach a maximum optical depth[27] of only $\tau \simeq 0.001$. We consider that the attenuation is due to Thomson scattering in the medium.

Therefore, the incident flux at a given distance $D$ from the Galactic center, taking into account also the attenuation of the radiation passing through the AGN torus and the interstellar medium, can be modeled as:

 

$$F_{\text{XUV}} = \frac{L_{\text{XUV}}}{4\pi D^2} e^{-\tau} \tag{1}$$

where $\tau$ is the optical depth along the line of sight.

**Atmospheric loss caused by XUV radiation.**　When XUV radiation falls upon a planetary atmosphere, it can cause atmosphere loss via hydrodynamic escape[16]. The energy-limited mass loss rate is usually modeled by a first-order differential equation[28]:

$$\dot{M} = \varepsilon \frac{3}{4} \frac{F_{\text{XUV}}}{G\rho_{\text{p}}} \tag{2}$$

where $F_{\text{xuv}}$ is the incident flux, $\rho_{\text{p}}$ is the planet density and $\varepsilon$ is an efficiency parameter that is estimated[29] in the range $0.1 \leq \varepsilon \leq 0.6$. The atmosphere mass lost over time for a given luminosity history of the AGN is obtained by integrating the above equation. Assuming an average luminosity $L_{\text{xuv}}$ over the AGN lifetime $\Delta t$, we can write the total mass lost by a planet at distance $D$ from the galactic center as:

$$M_{\text{lost}} = \frac{3\varepsilon}{12\pi} \frac{L_{\text{XUV}} \Delta t}{G\rho_{\text{p}} D^2} e^{-\tau} \tag{3}$$

As we showed earlier, the assumption of a lower Eddington ratio would provide the same integrated flux, but on a longer timescale. Therefore the total mass lost at the end of the AGN phase will not depend on the assumed luminosity. Moreover, we expect that the result will only depend marginally on the fact that the luminosity during the AGN phase is constant or flaring. Note also that the only relevant planetary parameter is the bulk density $\rho_{\text{p}}$, that we expect to be roughly the same of Earth for all rocky planets.

The previous equation can be inverted to obtain an analytic relation for the distance $D$ from the galactic center where a planet loses an atmosphere mass $M_{\text{lost}}$ after a time $\Delta t$:

$$D = \left[ e^{-\tau} \varepsilon \left( \frac{M_{\text{atm},\oplus}}{M_{\text{lost}}} \right) \left( \frac{L_{\text{XUV}}}{4.5 \times 10^{44}\,\text{erg}\,\text{s}^{-1}} \right) \left( \frac{\rho_{\oplus}}{\rho_{\text{p}}} \right) \left( \frac{\Delta t}{1.96 \times 10^7\,y} \right) \right]^{1/2} \text{kpc} \tag{4}$$

where $M_{\text{atm},\oplus} = 5 \times 10^{21}\,\text{g}$ is the mass of present day Earth's atmosphere and $\rho_{\oplus} = 5.5\,\text{g}\,\text{cm}^{-3}$ is the density of the Earth.

## Acknowledgements

The authors acknowledge useful conversations with Daniela Billi and Fausto Vagnetti. F.T. acknowledges support by the Programma per Giovani Ricercatori - 2014 "Rita Levi Montalcini".

## Author Contributions

A.B. had the original idea for this study and produced the numerical results, figures and most of the text. F.T. produced the estimates of the accretion history of the supermassive black hole, the AGN luminosity and the attenuation of XUV radiation, and wrote the corresponding sections. Both authors reviewed and discussed the full article.

## Additional Information

**Competing Interests:** The authors declare that they have no competing interests.

**Publisher's note:** Springer Nature remains neutral with regard to jurisdictional claims in published maps and institutional affiliations.